\documentclass[conference, 10pt]{IEEEtran}

\usepackage[hidelinks]{hyperref}
\usepackage{cite}
\usepackage{xcolor}
\usepackage{amsmath}
\usepackage{amssymb}
\usepackage{amsfonts}
\usepackage{graphicx}
\usepackage{textcomp, dirtytalk}
\usepackage{algorithm, algcompatible}
\usepackage{pgfplots}
\usepgfplotslibrary{groupplots}
\usetikzlibrary{matrix, fit}
\pgfplotsset{compat=1.17}

\IEEEtriggeratref{14}

\newcommand{\mat}[1]{\begin{pmatrix} #1 \end{pmatrix}}
\renewcommand{\vec}[1]{\mathbf{#1}}

\newcommand{\latencyBinaryMV}{39}
\newcommand{\latencyFullConv}{2}
\newcommand{\latencyBinaryConv}{12}

\begin{document}

\title{MatPIM: Accelerating Matrix Operations \\ with Memristive Stateful Logic}

% Double-blind
\author{
\IEEEauthorblockN{Orian Leitersdorf, Ronny Ronen, and Shahar Kvatinsky} \IEEEauthorblockA{\emph{Viterbi Faculty of Electrical and Computer Engineering, Technion -- Israel Institute of Technology, Haifa, Israel}} \IEEEauthorblockA{orianl@campus.technion.ac.il, ronny.ronen@technion.ac.il, shahar@ee.technion.ac.il}}

\IEEEoverridecommandlockouts

\IEEEpubid{\begin{minipage}{\textwidth}\ \\[12pt] \centering \copyright 2022 IEEE. Personal use of this material is permitted.  Permission from IEEE must be obtained for all other uses, in any current or future media, including reprinting/republishing this material for advertising or promotional purposes, creating new collective works, for resale or redistribution to servers or lists, or reuse of any copyrighted component of this work in other works.
\end{minipage}} 

\maketitle

% ---- Abstract ---- %
\begin{abstract}
The emerging memristive Memory Processing Unit (mMPU) overcomes the memory wall through memristive devices that unite storage and logic for real processing-in-memory (PIM) systems. At the core of the mMPU is stateful logic, which is accelerated with memristive partitions to enable logic with massive inherent parallelism within crossbar arrays. This paper vastly accelerates the fundamental operations of matrix-vector multiplication and convolution in the mMPU, with either full-precision or binary elements. These proposed algorithms establish an efficient foundation for large-scale mMPU applications such as neural-networks, image processing, and numerical methods. We overcome the inherent asymmetry limitation in the previous in-memory full-precision matrix-vector multiplication solutions by utilizing techniques from block matrix multiplication and reduction. We present the first fast in-memory binary matrix-vector multiplication algorithm by utilizing memristive partitions with a tree-based popcount reduction ($\mbox{\boldmath $\latencyBinaryMV\times$}$ faster than previous work). For convolution, we present a novel in-memory input-parallel concept which we utilize for a full-precision algorithm that overcomes the asymmetry limitation in convolution, while also improving latency ($\mbox{\boldmath $\latencyFullConv\times$}$ faster than previous work), and the first fast binary algorithm ($\mbox{\boldmath $\latencyBinaryConv\times$}$ faster than previous work). 
\end{abstract}

% ---- Keywords ---- %

% \vspace{-5pt}

\section{Introduction}
\label{sec:introduction}

Matrix operations construct the foundation for large-scale applications such as neural-networks, image processing, and numerical methods. Maximizing the efficiency of these operations has been thoroughly studied, such as Strassen's algorithm~\cite{Strassen}. We vastly accelerate matrix operations in the emerging memristive Memory Processing Unit (mMPU)~\cite{RealmMPU}.

\begin{figure}
    \centering
    \includegraphics[width=0.975\linewidth, trim={0cm 0.3cm 0cm 0cm}]{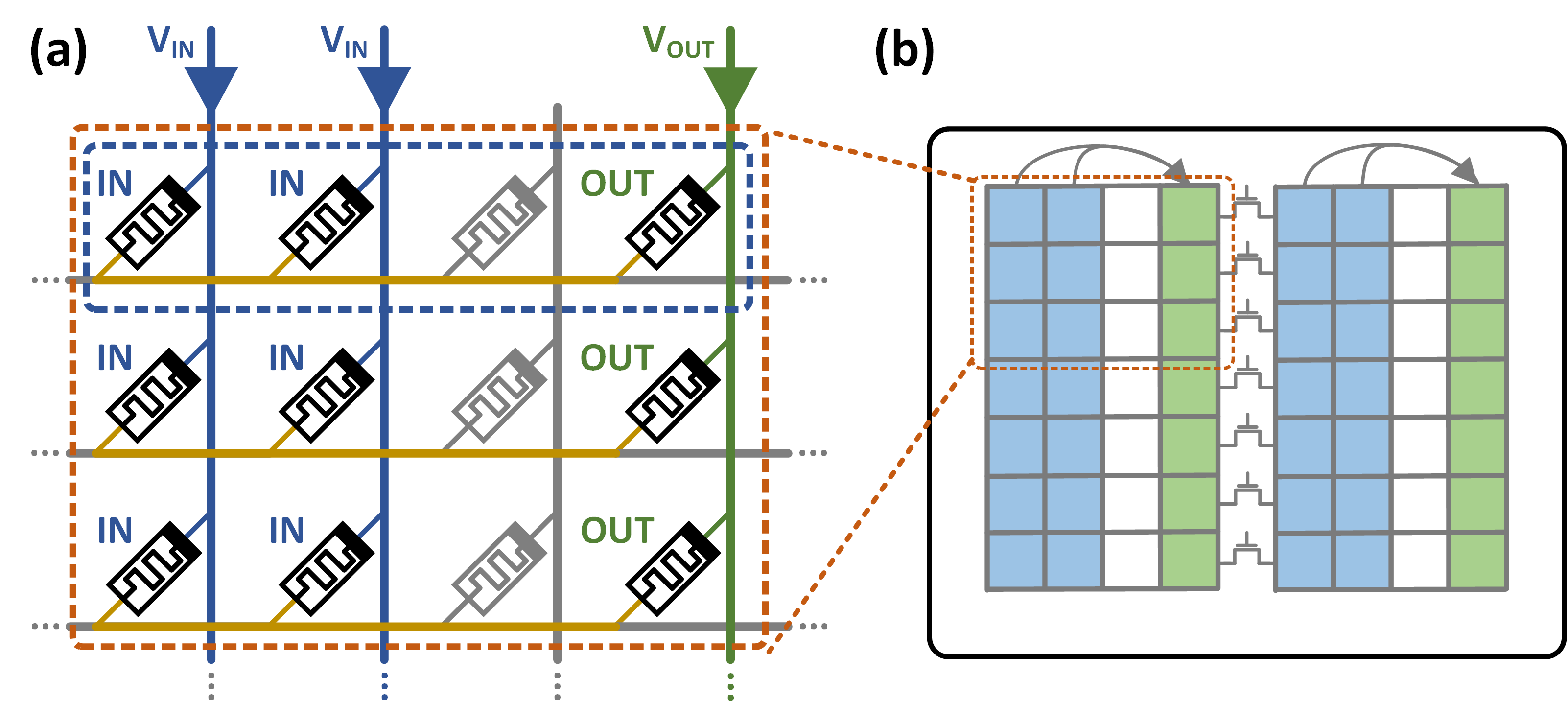}
    \caption{(a) Memristive stateful logic with parallelism across rows: bitwise logic on two columns into a third column in a single cycle. (b) Memristive partitions enable multiple concurrent column operations.}
    \label{fig:crossbar}
    \vspace{-15pt}
\end{figure}

The mMPU is rapidly emerging as a technology that may overcome the \emph{memory wall}~\cite{DarkMemory} through memristive~\cite{Memristor} crossbar arrays that enable real processing-in-memory (PIM)~\cite{NDP, MemristiveLogic, RACER}. At its core is the memristor~\cite{Memristor}, a two-terminal resistive device whose resistance may be modified with an applied voltage. This enables binary storage through the resistance value (low resistance for logical one, high resistance for logical zero), with memristive crossbar arrays essentially storing binary matrices. The experimentally-demonstrated~\cite{StatefulLogicReview, BarakVCM, LogicComputing} digital \emph{stateful logic}~\cite{MemristiveLogic} technique observes that applying voltages on bitlines/wordlines of memristive crossbars induces parallel logic within the crossbar, performed using the same memristors responsible for storage~\cite{IMPLY, MAGIC, FELIX}. For example, Figure~\ref{fig:crossbar}(a) demonstrates that applying voltages on bitlines induces a logic gate (e.g., NOR) in each row of the crossbar; essentially, a bit-wise operation on two columns is computed and stored in a third column within a single cycle. Single-row algorithms~\cite{RACER, MultPIM, ParityPIM, ICECS, abstractPIM, IMAGING, FloatPIM, Ameer, SIMPLER} utilize such parallelism for high-throughput vectored operation (e.g., vector addition) with latency independent of vector dimension: the arithmetic function (e.g., addition) is performed \emph{serially} within a single row (e.g., a single NOR at a time), yet repeated along all rows simultaneously. Memristive partitions~\cite{FELIX, RACER, MultPIM, alam2021sorting, ParityPIM, ICECS, RIME} dynamically divide the crossbar using transistors to enable multiple concurrent column operations, see Figure~\ref{fig:crossbar}(b), thereby enabling \emph{parallel} single-row algorithms that perform multiple concurrent gates within each row, across all rows~\cite{MultPIM}. 

\begin{figure*}
    \centering
    \includegraphics[width=\linewidth]{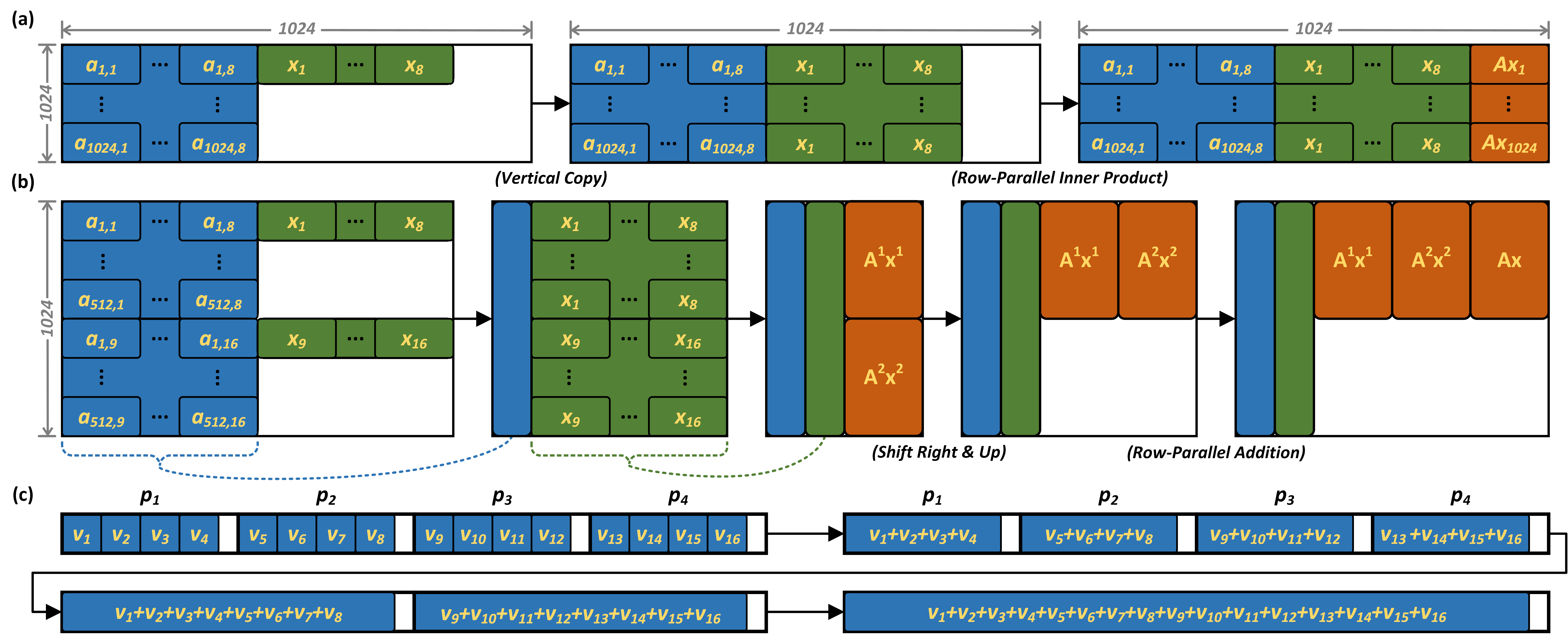}
    \caption{(a) The previous approach to full-precision in-memory matrix-vector multiplication, supporting only matrices of drastically non-symmetric dimension, e.g. $1024 \times 8$. (b) The proposed approach to in-memory matrix-vector multiplication that splits the computation using block matrix multiplication, supporting matrices of nearly any dimension, e.g. $64 \times 128$. (c) The proposed single-row approach to binary popcount, accelerated with memristive partitions.}
    \label{fig:matrixMult}
    \vspace{-5pt}
\end{figure*}

\IEEEpubidadjcol

We significantly advance in-memory matrix-vector multiplication~\cite{MultPIM, FloatPIM} and 2D convolution~\cite{FloatPIM, IMAGING}, operations at the core of many applications~\cite{numerical}. Matrices are stored within a single crossbar array and the goal is to compute and store the outputs in the crossbar through stateful operations. As we operate within a crossbar array using stateful logic, data-transfer is significantly reduced and high throughput follows from the concurrent operation of multiple crossbars~\cite{Bitlet}. We consider both full-precision (e.g., for traditional neural-networks) and binary-precision (e.g., for binary neural networks~\cite{XNORNet, BNNSurvey}) algorithms. Note that while memristive crossbar arrays can also be utilized in an analog fashion for fast matrix-vector multiplication~\cite{ProcIEEEAnalog}, we focus on approaches based on digital stateful logic. For the full-precision algorithms, this is justified due to the high precision (e.g., 32-bit) that is required in many applications, yet cannot be matched by an analog domain. For the binary algorithms, we avoid the throughput-limited sense amplifiers by processing data \emph{within} the crossbar; applications based on the proposed algorithms may process with high throughput by concatenating matrix operations within the same crossbar, thereby avoiding throughput-limited sense amplifiers.

Previously proposed matrix-vector\footnote{Matrix-matrix multiplication can be derived from repeated matrix-vector.} multiplication concepts~\cite{MultPIM, FloatPIM} utilize parallelism across rows/columns in a straightforward manner. Yet, for full-precision matrix multiplication, a critical asymmetry enables only matrix multiplication of highly-peculiar dimensions: $1024 \times 8$ (1024 rows but only 8 columns in the matrix) for $32$-bit integers and a crossbar sized $1024 \times 1024$~\cite{MultPIM}. We propose a technique based on block matrix multiplication and reduction that overcomes this asymmetry, e.g., enabling dimension $64 \times 128$. We also present the first fast binary matrix-multiplication technique using a tree-based reduction with partitions, which is $\latencyBinaryMV\times$ faster than the special case of $1$-bit in previous approaches~\cite{MultPIM, FloatPIM}. 

Previous approaches for in-memory convolution either require non-trivial periphery~\cite{FloatPIM} or suffer from the same asymmetry above~\cite{IMAGING, FloatPIM}. We propose a novel in-memory approach to convolution, enabling fast efficient operation for both full and binary precision. The proposed full-precision algorithm both overcomes the asymmetry limitation and even improves latency by $\latencyFullConv\times$ over previous work~\cite{IMAGING}, due to more efficient shifting. The binary algorithm improves latency by $\latencyBinaryConv\times$ over the case of $1$-bit integers in previous work~\cite{IMAGING}. 

\section{In-Memory Matrix-Vector Multiplication}
\label{sec:mult}

This section presents fast in-memory matrix-vector multiplication, outperforming the previous state-of-the-art~\cite{FloatPIM, MultPIM}. 

\subsection{Full-Precision: Balanced Matrix-Vector Multiplication}
\label{sec:mult:full}

We address the asymmetry in in-memory matrix multiplication, enabling fast full-precision matrix multiplication with flexible dimensions. Let matrix $\vec{A}$ be of dimensions $m \times n$ and vector $\vec{x}$ of dimension $n$ be stored in the memory, the goal is to compute and store $\vec{Ax}$ within the crossbar, where each element in $\vec{A}, \vec{x},$ and $\vec{Ax}$ is an $N$-bit number. 

\begin{figure*}
    \centering
    \includegraphics[width=0.95\linewidth, trim={0cm 1.8cm 0cm 0cm}]{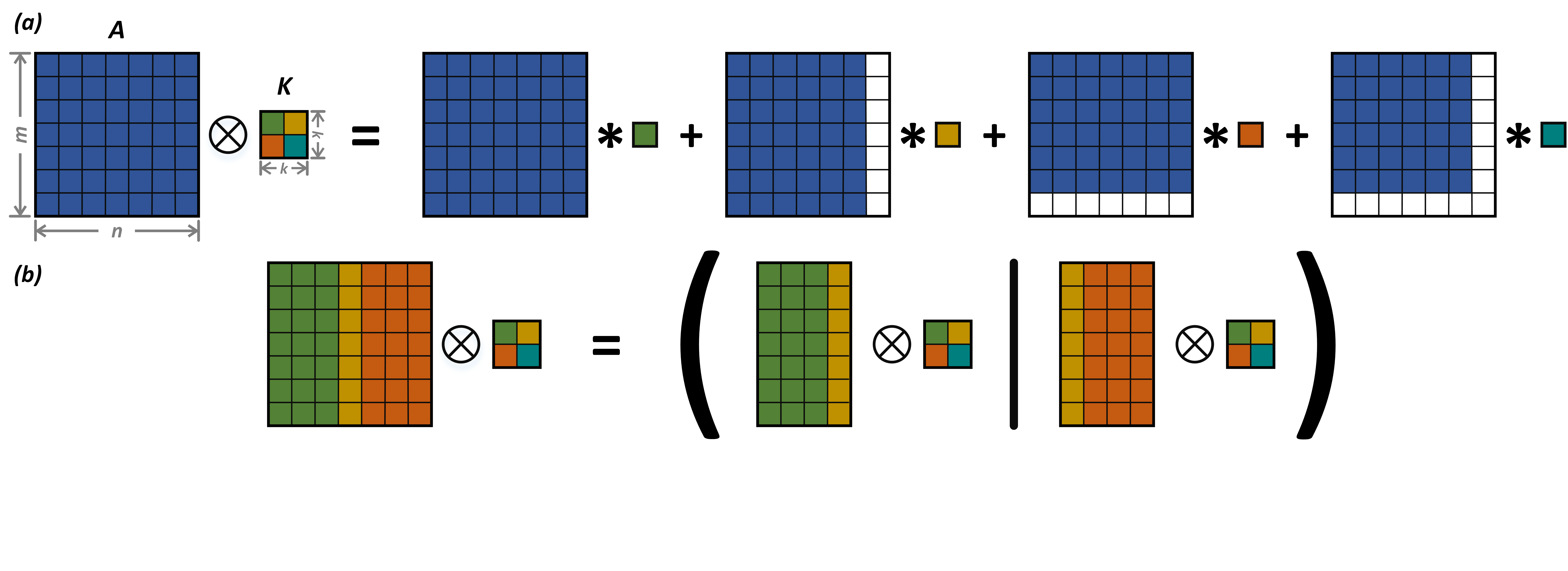}
    \caption{(a) Input-parallel convolution concept which computes the sum of a shifted versions of $\vec{A}$ multiplied by single elements from $\vec{K}$. (b) Convolution property which enables splitting the input $\vec{A}$ into overlapping blocks $\vec{A}^1, ..., \vec{A}^\alpha$ and computing $\vec{A} \otimes \vec{K}$ by concatenating the results of $\vec{A}^1 \otimes \vec{K}, ..., \vec{A}^\alpha \otimes \vec{K}$.}
    \label{fig:matrixConv}
    \vspace{-5pt}
\end{figure*}

The basic previous concept is shown in Figure~\ref{fig:matrixMult}(a), storing each $N$-bit element horizontally. Vector $\vec{x}$ is duplicated to rows with stateful operations across rows, and then each row performs an inner-product (multiply and accumulate) between a row of $\vec{A}$ and $\vec{x}$. The inner product is performed in parallel across rows, using column operations~\cite{MultPIM, FloatPIM}. The overall latency is $O(m+nN\log N)$, due to (1) duplicating $\vec{x}$ in $O(m)$ cycles, and (2) multiplying two $N$-bit numbers within a single row in $O(N \log N)$ cycles, which is repeated $n$ times~\cite{MultPIM}. While this elegantly utilizes parallelism, scalability is limited: as elements are stored horizontally ($1 \times N$ memristors), a $1024\times 1024$ crossbar can support $m=1024$ but only $n \leq 8$, for $N=32$. This is problematic, as matrix-vector multiplication of dimension $1024 \times 8$ is highly uncommon.

Our proposed concept overcomes this asymmetry by splitting the matrix-vector multiplication into blocks. The concept follows from block matrix multiplication: consider splitting $\vec{A}$ to left and right halves, and $\vec{x}$ to top and bottom halves,
\begin{equation}
    \vec{A} = \left(\vec{A}^1 \; \vec{A}^2\right), \vec{x} = \mat{\vec{x}^1 \\ \vec{x}^2} \implies \vec{Ax}  = \vec{A}^1\vec{x}^1 + \vec{A}^2\vec{x}^2.
    \label{eq:mult:proposed}
\end{equation}
We generalize (\ref{eq:mult:proposed}) to $\alpha$ blocks. Our algorithm begins by computing each pair $\vec{A}^i\vec{x}^i$ across its own $m$ rows (taking $\alpha m$ rows in total). Thus, the computation of all $\alpha$ pairs is performed in parallel. We continue by summing the vectors ($\vec{A}^1\vec{x}^1, \vec{A}^2\vec{x}^2,...,\vec{A}^\alpha \vec{x}^\alpha$) through a logarithmic technique inspired by reduction~\cite{Bitlet}. We start with $\alpha$ vectors stored vertically, shift half of them to the right and upwards, and add the vectors in parallel, reducing the task from $\alpha$ vectors to $\alpha/2$. Then, we continue recursively. Figure~\ref{fig:matrixMult}(b) demonstrates the case of $\alpha=2$. Overall, we attain a latency of $O(\alpha m + (n/\alpha) N \log N + N\log\alpha + \alpha m)$, due to (1) the initial copying, (2) the parallel inner products, and (3) the reduction.

\subsection{Binary: Fast Matrix-Vector Multiplication}
\label{sec:mult:binary}

Binary matrix-vector multiplication is where the elements in $\vec{A}$ and $\vec{x}$ are binary (e.g., $-1$ or $1$~\cite{XNORNet, BNNSurvey}), and the output is quantized (e.g., majority). The naive application of the previous approach~\cite{FloatPIM, MultPIM} performs the same algorithm for full-precision multiplication even for $N=1$. The inner product in each row is performed by multiplying the $1$-bit numbers (e.g., XNOR~\cite{XNORNet, BNNSurvey}), adding each product to a counter, and then comparing the sum to $n/2$ (majority). In Section~\ref{sec:evaluation}, we show this is highly inefficient.

As this summation is essentially a popcount operation, we propose two improvements that vastly improve performance.
First, popcount can be implemented more efficiently with a tree rather than a counter, due to the representation size increasing throughout the summation. Second, using partitions, we accelerate the single-row popcount with a tree-based reduction. Assuming $p$ partitions and a pop-count of $n$ bits, we start with each partition storing $n/p$ bits. All partitions start with a serial popcount operation on their $n/p$ bits, in parallel across all partitions. We continue with a reduction tree amongst the partitions. For example, the first recursive step would connect every pair of partitions (set transistor between them to conducting) and add in parallel across $p/2$ partitions. Figure~\ref{fig:matrixMult}(c) illustrates the case of $n=16$ and $p=4$.

\section{In-Memory Convolution}
\label{sec:conv}

In-memory convolution is more complex than matrix-vector multiplication due to the required shifting~\cite{IMAGING, FloatPIM}. Thus, convolution in FloatPIM~\cite{FloatPIM} requires barrel-shifters for each crossbar array, incurring undesirable area and control overhead~\cite{RACER}. Conversely, IMAGING~\cite{IMAGING} does not utilize barrel shifters, yet requires a complex output-parallel technique. We present a simple input-parallel technique that improves both of the previous methods, with both full-precision and binary algorithms. Consider the 2D convolution $\vec{A} \otimes \vec{K}$ of input $\vec{A}$ (dimension $m \times n$) and kernel $\vec{K}$ (dimension $k \times k$). 

\subsection{In-Memory Input-Parallel Convolution}
\label{sec:conv:inputParallel}

\begin{algorithm}[t]
 \small
 \caption{In-Memory Input-Parallel Convolution}
 \begin{algorithmic}[1]
 \renewcommand{\algorithmicrequire}{\textbf{Input:}}
 \renewcommand{\algorithmicensure}{\textbf{Output:}}
 \REQUIRE Input $\vec{A}$ (dim. $m \times n$), Kernel $\vec{K}$ (dim. $k \times k$)
 \ENSURE $\vec{A} \otimes \vec{K}$, with no padding (w.l.o.g.)
 
 \STATE $\vec{Out} \gets \vec{0}_{(m-\lceil k/2\rceil) \times (n-\lceil k/2 \rceil)}$
 
 \FOR {$vert = 0$ to $k-1$}
 \FOR {$hori = 0$ to $k-1$}
 \FOR {$col = 0$ to $n-\lceil k/2 \rceil - 1$}
 \STATE // Row-parallel addition/multiplication of columns
 \STATE $\vec{Out}[:, col]  \mathrel{{+}{=}} \vec{A}[:, col + hori] * \vec{K}[vert][hori]$
 \ENDFOR
 \ENDFOR
 \STATE \text{Shift $\vec{A}$ vertically once (upwards).}
 \ENDFOR
 \STATE \textbf{return} $\vec{Out}$
 \end{algorithmic} 
 \label{alg:inputParallelConvolution}
 \end{algorithm}

IMAGING considered an output-parallel approach: for each element in the output matrix, they compute all of the products for that element in parallel; yet, this led to complex expensive movements between multiplications. Conversely, we propose an input-parallel approach: for each element $a$ in the input $\vec{A}$, we consider the elements in $\vec{A}\otimes\vec{K}$ that $a$ contributes to. We find that $a$ is multiplied only once by each element in $\vec{K}$ for each of the neighbors of that element in $\vec{A} \otimes \vec{K}$. Thus, our approach constructs $\vec{A} \otimes \vec{K}$ from the sum of shifted versions of $\vec{A}$ multiplied by single elements from $\vec{K}$ (input-parallel), as shown in Figure~\ref{fig:matrixConv}(a).
The in-memory implementation of this concept is described in Algorithm~\ref{alg:inputParallelConvolution}: we initialize a zero output matrix, and for each element $x$ in $\vec{K}$, we add $\vec{A} * x$ to the current sum and then shift $\vec{A}$ to match the next element. Specifically, the multiplication is performed by duplicating $x$ across all rows and then multiplying columns from $\vec{A}$ with $x$ in parallel. The horizontal shifts are simulated as part of the access (similar to IMAGING), and the vertical shifts are performed with stateful gates alone (no barrel shifter) along rows. 
The barrel shifters from FloatPIM are not required since the shift is performed in parallel across the entire row (due to the input-parallel approach), and thus the latency of naive shift based on stateful logic is amortized. This shift parallelism is also the benefit of the proposed approach over IMAGING.

\subsection{Balanced Full-Precision Convolution}
\label{sec:conv:full}

Similar to matrix multiplication, the proposed concept in Section~\ref{sec:conv:inputParallel} for in-memory matrix convolution suffers from asymmetry in full-precision form, thus in this section a balanced implementation that overcomes that limitation is proposed. Figure~\ref{fig:matrixConv}(b) shows a simple property of convolutions that enables splitting $\vec{A}$ into multiple overlapping blocks, $\vec{A}^1, ..., \vec{A}^\alpha$, and computing the result $\vec{A} \otimes \vec{K}$ based on the concatenation of $\vec{A}^1 \otimes \vec{K}, ..., \vec{A}^\alpha \otimes \vec{K}$ ($\alpha = 2$ in the illustration). Therefore, while $\vec{A}$ may not fit in a crossbar when each element in $\vec{A}$ is stored horizontally ($1 \times N$) as the row size is the bottleneck, a block $\vec{A}^i$ can fit (as it contains less columns). Thus, we vertically concatenate the blocks $\vec{A}^1, ..., \vec{A}^\alpha$ within a single crossbar array, in a manner similar to the balanced matrix-vector multiplication. Note that the convolution occurs in parallel across all blocks as they are stacked vertically.

\begin{table}[t]
    \centering
    \caption{Matrix-Vector Multiplication Latency [Cycles]}
    \begin{tabular}{c|c|c|c|c}
        $\vec{A}$ & $\vec{x}$ & $N$ & Baseline~\cite{MultPIM, FloatPIM} & Proposed \\
         \hline\hline
        $1024 \times 8$ & $8 \times 1$ & 32 & 4657 & 4657 \\
        $512 \times 16$ & $16 \times 1$ & 32 & Not Supported & 5367 \\
        $256 \times 32$ & $32 \times 1$ & 32 & Not Supported & 5822 \\
        $128 \times 64$ & $64 \times 1$ & 32 & Not Supported & 6151 \\
        \hline\hline
        $1024 \times 384$ & $384 \times 1$ & 1 & 14770 & 383 \\
        % (1*1+2*2+4*3+8*4+16*5+32*6+64*7+128*8+(384-255)*9)*5 = 14770
    \end{tabular}
    \label{tab:multResults}
    \vspace{-5pt}
\end{table}

\subsection{Fast Binary Convolution}
\label{sec:conv:binary}

We propose fast binary convolution based on the concept from Section~\ref{sec:conv:inputParallel} and the technique for splitting convolution introduced in Section~\ref{sec:conv:full}. Essentially, we compute each $\vec{A}_i \otimes \vec{K}$ in its own memristive partition. This case is simpler than the binary matrix-vector multiplication as each inner product fits within a single partition and thus no inter-partition communication is necessary. Furthermore, we replace the full-precision multiply and accumulate used in the full-precision convolution with the single-row single-partition popcount operation introduced in Section~\ref{sec:mult:binary}.

\section{Evaluation}
\label{sec:evaluation}

We evaluate the proposed algorithms for full-precision and binary matrix-vector multiplication and convolution, demonstrating significant improvement as compared to the previous in-memory stateful-logic works in terms of flexibility with dimensions (full-precision) and latency (full-precision and binary). Results are verified via a custom cycle-accurate simulator\footnote{Available at https://github.com/oleitersdorf/MatPIM.} that models the logic operations in the array, with the proposed algorithms executing a sequence of stateful operations and the simulation environment verifying correctness. We choose a crossbar that supports the FELIX~\cite{FELIX} suite of logic gates, and modify the results from previous works~\cite{IMAGING, FloatPIM} to assume the state-of-the-art arithmetic for addition and multiplication~\cite{MultPIM}, providing a fair comparison of the algorithmic concepts rather than the specific stateful-logic technique. We consider a $1024 \times 1024$ crossbar array with $32$ partitions within rows and columns.

\begin{table}[t]
    \centering
    \caption{2D Convolution Latency [Cycles]}
    \begin{tabular}{c|c|c|c|c}
        $\vec{A}$ & $\vec{K}$ & $N$ & Baseline~\cite{IMAGING} & Proposed \\
         \hline\hline
         $1024 \times 4$ & $3 \times 3$ & 32 & 28760 & 15352 \\
         \hline
        %  $340 \times 6$ & $3 \times 3$ & 32 & 10964 & \\
         $1024 \times 8$ & $3 \times 3$ & 32 & Not Supported & 39897 \\
         $512 \times 16$ & $3 \times 3$ & 32 & Not Supported & 49092 \\
         $256 \times 32$ & $3 \times 3$ & 32 & Not Supported & 49592 \\
         $128 \times 64$ & $3 \times 3$ & 32 & Not Supported & 49824 \\
         \hline
        $1024 \times 8$ & $5 \times 5$ & 32 & Not Supported & 81305 \\
        $512 \times 16$ & $5 \times 5$ & 32 & Not Supported & 127728 \\
        $256 \times 32$ & $5 \times 5$ & 32 & Not Supported & 128220 \\
        $128 \times 64$ & $5 \times 5$ & 32 & Not Supported & 128436 \\
        \hline\hline
        $1024 \times 256$ & $3 \times 3$ & 1 & 45312 & 3805 \\ % 3*(2*1024+256*3*4+256*5+256+2*1024+10*256+15*256) = 45312
    \end{tabular}
    \label{tab:convResults}
    \vspace{-5pt}
\end{table}

\subsection{Matrix Multiplication}

Table~\ref{tab:multResults} summarizes the results for in-memory matrix multiplication, for both the full-precision ($N=32$)\footnote{$N=32$ is chosen as an common example, we also support different $N$.} and binary ($N=1$) algorithms. The proposed full-precision algorithm supports far greater dimension flexibility than the previous works~\cite{MultPIM, FloatPIM}; for example, the previous works only supported $1024 \times 8$, while we support $512 \times 16, 256 \times 32, 128 \times 64$. For the binary algorithms, we present a substantial improvement of $\latencyBinaryMV\times$ due to the combination of two techniques: optimized popcount and partition-based reduction tree.

\subsection{Matrix Convolution}

Table~\ref{tab:convResults} summarizes the results for in-memory 2D convolution, for both the full-precision ($N=32$) and binary ($N=1$) algorithms. The proposed full-precision algorithm both possess greater dimension flexibility than the previous work~\cite{IMAGING} and improves latency by $\latencyFullConv\times$. In the binary case, we present a drastic improvement of $\latencyBinaryConv\times$ due to the combination of the in-memory input-parallel concept and the distributed division of the convolution amongst partitions.

% \section{Related Works}
% \label{sec:related}

% While this paper considered in-memory matrix-multiplication and convolution based on digital stateful logic, there exists a parallel field investigating an efficient near-memory analog matrix-vector multiplication in memristive crossbars~\cite{ProcIEEEAnalog}. Yet, such solutions typically require high-overhead periphery for the conversion between the analog and digital domain, and transfer the data outside the crossbar array (limiting throughput). Conversely, the proposed solutions operate solely in the digital domain, and avoid the throughput-limited data-transfer by operating completely within the crossbar using high-throughput stateful operations.

\section{Conclusion}
\label{sec:conclusion}

This paper vastly accelerates matrix-vector multiplication and convolution for the memristive Memory Processing Unit (mMPU). For matrix-vector multiplication, we extend the previous concept with block matrix multiplication to overcome an asymmetry challenge in full-precision algorithms, and with partitions to provide fast binary algorithms. For convolution, we propose an in-memory input-parallel approach, reducing the requirement for shift operations and not requiring an external barrel shifter; this provides significant improvement for full-precision algorithms and enables fast binary convolution. While alternative techniques such as analog memristive computing are widely explored for neural networks, the mMPU is also an attractive architecture for data-intensive applications due to the high-throughput \emph{within} crossbar arrays and low area overhead; the expansion of the proposed algorithms to applications such as neural networks will be investigated in future work. Overall, we provide an efficient foundation for the operations at the core of many applications, thereby advancing the mMPU towards a new era of in-memory computing.

\section*{Acknowledgment}
This work was supported in part by the European Research Council through the European Union's Horizon 2020 Research and Innovation Programe under Grant 757259, and in part by the Israel Science Foundation under Grant 1514/17.

% ---- References ---- %
\bibliographystyle{IEEEtran}
\bibliography{refs}

\end{document}